\documentstyle[12pt,aasms4]{article}        

\def\cn{SN~1997cn}
\def\bg{SN~1991bg}
\def\k{SN~1992K}
\def\kms{km s$^{-1}$}

\def\Lsun{L_{\odot}}
\def\M{$M_{\odot}$}

\def\Teff{T_{\ast}}
\def\m100{mag/100$^d$}
\def\ni{{$^{56}$Ni}}
\def\c57{{$^{57}$Co}\/}

\def\ti44{{$^{44}$Ti}\/}
\def\r0{{$R_0$}}
\def\ltsima{$\; \buildrel < \over \sim \;$}
\def\ltsim{\lower.5ex\hbox{\ltsima}}
\def\gtsima{$\; \buildrel > \over \sim \;$}
\def\gtsim{\lower.5ex\hbox{\gtsima}}

\begin{document}
\lefthead{Turatto et al.}
\righthead{SN~1997cn} 

\title{A new faint type Ia supernova: SN~1997cn in NGC 5490 
\footnote[0]{Based on observations collected at ESO-La Silla}}

\author{M. Turatto\altaffilmark{1}, 
A. Piemonte\altaffilmark{2}, 
S. Benetti\altaffilmark{3,4}, 
E. Cappellaro\altaffilmark{1},
P.A. Mazzali\altaffilmark{5}
I.J. Danziger\altaffilmark{5},
F. Patat \altaffilmark{4}    }

\altaffiltext{1}{Osservatorio Astronomico di Padova, vicolo 
dell'Osservatorio 5, I-35122 Padova, Italy}
\altaffiltext{2}{Departamento de Astronomia, 
P. Universitad Catolica, Casilla 104, Santiago 22, Chile}
\altaffiltext{3}{Telescopio Nazionale Galileo, Apartado de Correos 565, 
E-38700 Santa Cruz de La Palma, Canary Islands, Spain}
\altaffiltext{4}{European Southern Observatory, Alonso de Cordova
3107, Vitacura, Casilla 19001 Santiago 19, Chile}
\altaffiltext{5}{Osservatorio Astronomico di Trieste, via G.B.
Tiepolo 11, I-34131 Trieste, Italy}

\begin{abstract}
Observations of the recent \cn\/ in the elliptical galaxy NGC~5490
show that this objects closely resembles, both photometrically and
spectroscopically, the faint SN~Ia \bg\/.  

The two objects have similar light curves, which do not show secondary
maxima in the near IR as normal type Ia supernovae.  The host galaxy,
NGC~5490, lies in the Hubble flow. Adopting for \cn\/ a reddening
$E(B-V)=0$, the absolute magnitude is faint: $M_V = -17.98$ using
$H_o=65$ and $M_V = -17.40$ using $H_o=85$ \kms Mpc$^{-1}$.  The
latter value is in close agreement with the absolute magnitude of
\bg\/ on the SBF--PNLF--TF distance scale.  The photospheric spectra
of the two SNe show the same peculiarities, the deep
\ion{Ti}{2} trough between 4000 and 4500A, the strong \ion{Ca}{2} IR
triplet, the narrow absorption at about 5700A\/ and the slow expansion
velocity.

In analogy to \bg\/ the observed spectrum of \cn\/ has been successfully 
modeled by scaling down the W7 model by a factor of 2, assuming a
rise time to B maximum of 18 days, a photospheric velocity and an effective
temperature low compared to normal SNIa. The influence
of the distance scale adopted on the input parameters of the best fit
model is also discussed.

These data demonstrate that peculiar SNIa like \bg\/ are not
once--in--a--lifetime events and that deep SN searches can be
contaminated by underluminous SN~Ia in a fairly large volume.

\end{abstract}


\section{Introduction}

Type Ia SNe have been used for decades as standard candles up to
cosmological distances because of their high luminosity and apparent
homogeneity. However, starting from the late eighties, this latter
property has been challenged (e.g. \cite{bran87}).
In 1991 two objects were
discovered, the faint \bg\/ (\cite{fil92a}, \cite{leib93},
\cite{tur91bg}) and the bright SN~1991T (\cite{fil92b},
\cite{polen92}, \cite{phil92}), which unambiguously demonstrated that SNe Ia
experience a wide range in luminosities, expansion velocities and
effective temperatures. More recently, accurate analyses of recent
high quality data (e.g. \cite{phil93}, \cite{riess96}) have shown that
the diversity among SNe Ia is not limited to some peculiar cases, and
that even "normal" SNe Ia cover a range of properties at either end of
which are the two extreme objects mentioned above.

The observations at both early and late epochs seem to
require a range of one order of magnitude both in the mass of
radioactive material and in the total mass of the exploding stars
(\cite{cap97}). This is reflected in the kinetic energy of the ejecta
(\cite{maz98}). Hence the standard scenario for SNe Ia progenitors,
i.e. a white dwarf accreting mass up to the Chandrasekhar limit, has
been challenged, and new progenitors are being investigated
(e.g. \cite{hoflkhok}).  In these models, variable amounts of heavy
elements are produced and returned to the ISM via the explosion. Since
SNe Ia may have a diversity of progenitors, the current models of
galaxy chemical evolution should be revised. Obviously, the observed
diversity has also major implications for the use of SNe Ia as distance
indicators.

Although we are now aware of the diversity within SNe Ia, we do not
know yet the distribution of SNe Ia with respect to these properties.
Are extreme cases like \bg\/ and SN~1991T just rare events, which do
not affect the average properties of the overwhelming majority of
"normal" SNe Ia? Are they distinct sub-types, or do they represent the
extremes of a continuous distribution of properties?  Are
intrinsically faint events severely underestimated, to the extent that
they may be so frequent as to constitute the majority of SNe Ia?

So far only 4 intrinsically dim SNe Ia have been discovered:
the prototype \bg; SN 1986G (\cite{phil87}, \cite{cris92}); SN 1992K
(\cite{ham94}); SN 1991F (\cite{gomlop}). However, the last 3 did not share
the same extreme properties as SN 1991bg (cfr. \cite{tur91bg} and
references therein).

In this paper we present the observations of a new candidate member
(\cite{turiau}) of the class of faint SNe Ia, and discuss its
properties.  \cn\/ was discovered on May 14.6 UT (\cite{li}) 6.7
arcsec East and 11.7 arcsec South of the nucleus of the relatively
distant elliptical galaxy NGC 5490 ($V_{3K}=5246$ $km s^{-1}$; RC3).

\begin{table}
\dummytable\label{phlog}
\end{table}

\section{Observations}

The data presented in this paper were collected at ESO, La Silla with
different telescopes (cf. Tables~\ref{phlog} and \ref{splog}) starting
soon after discovery and until the object became too faint for our
instrumentation.  

The SN photometry is given in Table~\ref{phlog}. The photometry was
obtained with respect to a local sequence (Fig.~\ref{snfield})
calibrated with respect to the photometric sequence around Nova
Sakurai (\cite{duersb}), which was observed in three photometric
nights together with the SN field. Table~\ref{seq} gives the
magnitudes obtained for the SN local sequence and the internal errors
for stars measured in different nights.  In order to subtract the galaxy
contribution, the PSF-fitting procedures implemented in the Romafot
package were used.  The uncertainty in the SN photometry varied with
time because of the decreasing contrast of the SN against the galaxy
background.
Estimates of the errors ranging from 0.05 mag at early epochs to 0.2
mag in the later V observations, are reported in brackets.

The spectra were calibrated using a standard technique, including flat
fielding, optimal extraction, wavelength and flux
calibration with respect to spectrophotometric standard stars observed
on the same nights.

\begin{table}
\dummytable\label{seq}
\end{table}

\begin{table}
\dummytable\label{splog}
\end{table}

In Fig.~\ref{figlc} we show the light curves of \cn\/ in UBVRI.
Clearly, the monitoring started when the SN had just passed maximum
light. However, the templates plotted in Fig.~\ref{figlc} show that
the light curve of \cn\/ resembles closely in shape that of \bg. Thus
we can confidently estimate the epochs and the magnitudes of the
maxima in the various bands (cf. Tab.~\ref{sum}).  The discovery
magnitude (``mag about 15.8 on an unfiltered CCD frame'';IAUC 6661) is
somewhat inconsistent with the rest of the light curve, but we choose
to ignore it because it is not in a defined photometric band. At the
epoch of the last observation the SN was not detected. Unfortunately
the photometric detection limit is not deep enough to establish the
late decline rate.


As with \bg, \cn\/ faded monotonically in R and I, showing no
sign of the secondary maxima characteristic of the light curves of
"normal" SNe Ia in the near IR. A comparison with SN~1994D
(\cite{pat}) in Fig.~\ref{complc} illustrates this difference.  Also,
the (B--V) colour curve of \cn\/ is different from that of "normal"
SNe Ia and similar to, although slightly bluer than that of \bg\/ (cf.
Fig.~\ref{complc}).

\section{Data analysis and modeling}

Phillips (1993) showed that the rate of decline of the light curve
after maximum light correlates with the absolute magnitude at
maximum. This was confirmed by Hamuy et al. (1996) for a larger sample
of SNe. Riess et al. (1996) extended the treatment to show that the
whole shape of the light curve depends on the absolute magnitude at
maximum. Since the light curve of SN~1997cn is very similar to that of
SN~1991bg, we can use the above-mentioned correlations to predict that the
absolute magnitude of SN~1997cn is faint.

To quantify this indication we can use both the relation between the
peak luminosity and $\Delta m_{15}$ (\cite{phil93,ham96}) and the
multicolour light curve shape (MLCS) technique (\cite{riess96}). The
MLCS method is based on the comparison between the BVRI light curves
of a SN and those of well-observed objects defining a training
set. Applying this method to \cn\/ we obtain a value of the
"luminosity correction" $\Delta = 1.50\pm0.05$. This should be
compared to values $\Delta = 1.44$ for \bg\/ and $\Delta = 1.25$ for
\k\/ (\cite{riess96}). Taken at face value, the MLCS method indicates
a ``negative'' reddening E(B--V)$=-0.10$. This simply tells us that
\cn\/ was bluer than \bg, the object used to define the behaviour of
faint SNe Ia (\cite{riess96}).  In that work it was assumed that
E(B--V)=0 for \bg\/, while some small reddening was probably present
(E(B--V)=0.05; \cite{tur91bg}). Considering that the parent galaxy is
elliptical, that galactic extinction is A$_{\rm B} = 0.0$
(\cite{burheil}), and the MLCS indication, we conclude that \cn\/ did
not suffer from any reddening along the line of sight.

\begin{table}
\dummytable\label{sum}
\end{table}

To estimate the distance modulus of NGC5490 we must rely on the
recession velocity and adopt a value for the Hubble constant.  
A value of $H_o =85$ \kms Mpc$^{-1}$ gives $m-M=33.95$, whereas
$H_o = 65$ implies $m-M=34.53$.
Based on the observed magnitude at
maximum, we obtain $M_V = -17.40$ ($H_o=85$) or $M_V = -17.98$
($H_o=65$). As we will discuss later, the different luminosity
calibrations correspond to different solutions for the spectrum
synthesis model.

Regardless of the adopted distance scale the SN was quite faint.
Indeed the fainter value is in close agreement with the absolute
magnitude of SN~1991bg, $M_V = -17.28$ (\cite{tur91bg}). The magnitude
of SN~1991bg which, derived using the SBF--PNLF--TF distance
scale (e.g. \cite{vaug95}), is consistent with $H_0 = 85$.

The rate of decline in the first 15 days after maximum is $\Delta
m_{15}(B) = 1.86$. This value is very close to that of \bg\/ ($\Delta
m_{15}(B) = 1.95$, \cite{tur91bg}), suggesting that the two SNe
reached about the same luminosity at maximum. One can notice that
\bg\ and \cn\ deviate from the linear relation between $\Delta m_{15}$
and absolute magnitude as defined from ``average'' SN~Ia
(\cite{ham96}). The other faint SN~Ia SN~1992K seems to conform to
this relation but since the first observations of SN1992K were
obtained only 10 days after maximum the value of $\Delta m_{15}$ is
quite uncertain.

The spectral evolution of \cn\/ is outlined in Fig.~\ref{compsp},
where we have compared the spectra of \cn\/ and \bg\/ at two different
epochs. The maximum light spectrum of \cn, which was used for the
early SN classification (\cite{turiau}), is different from those of
normal SNe Ia, and resembles closely that of \bg.  In particular this
is true in the region between 4200 and 4500A, where both SNe show a
broad absorption due to \ion{Ti}{2} (\cite{fil92a},
\cite{maz97}). Also, the profile of the \ion{Si}{2} 6355A\/ line
is the same, indicating a very slow expansion velocity of the
ejecta. The only notable difference is in the intensity of the
emission at 4600A\/ which is stronger in \cn\/ than in \bg.

For the maximum epoch we have computed the line ratio $\Re$(\ion{Si}{2})
(\cite{nug95}), which defines a spectroscopic sequence for SNe Ia with
values ranging from $\Re=0.14$ for SN~1991T to $\Re=0.62$ for \bg.
For \cn\/ we measured $\Re$(\ion{Si}{2})=0.63, which is very similar
to the value for \bg, as expected.  The value of $\Re$(\ion{Ca}{2}) is
rather uncertain for \cn\/ due to the poor signal--to--noise ratio of
our spectrum at this wavelength.

The second comparison is between \cn\/ and \bg\/ using spectra taken
about one month after maximum.  The spectra are again very similar,
with the same line intensity ratios and widths. The difference in the
intensity of the continuum below 5000A\ is probably due to a poor
galaxy background subtraction in the noisy spectrum of \cn.  The
\ion{Ca}{2} IR triplet is very strong in both objects compared to
other SNe Ia (cfr. Fig.~8 of \cite{tur91bg}). The very narrow absorption
at about 5700A, which was one of the most mysterious features of
\bg\/ (\cite{tur91bg}), is visible in \cn\ as well.

We also obtained a spectrum of \cn\/ 2.5 months after
discovery. Although the spectrum is very noisy, it does resemble the
spectrum of \bg\/ at 79 days after maximum. In particular, the
[\ion{Ca}{2}] 7291,7324 A\/ emission is very strong. This
line is also the dominant feature in the nebular spectra of SN 1991bg.

We have modeled the maximum light spectrum of \cn\/ using the new
version of the Monte Carlo spectrum synthesis code by Mazzali \& Lucy
(1998, in preparation), which includes photon branching.

Since \cn\/ and \bg\/ are so similar, we fitted \cn\/ using parameters
similar to those used to fit the maximum light spectrum of \bg\/
(\cite{maz97}). Those results are summarized here. The best fit for \bg\/
was obtained using $\mu=31.13$ for NGC4374 which corresponds to a
short distance scale. We used $E(B-V)=0.05$ for the reddening and
obtained a best fit by assuming that the exploding mass was only 0.7
\M, derived by scaling the W7 density structure (\cite{nomoto})
down by a factor of 2.  Maximum light was assumed to occur 18 days
after the explosion.  The best-fitting model had luminosity $\log L /
\Lsun = 42.33$ erg s$^{-1}$, photospheric velocity $v_{ph}=6750$ km
s$^{-1}$, so that the effective temperature was $\Teff = 7050$K. The
abundances were ad-hoc, that is compared to W7 O was increased, the
Fe-group elements decreased, and among the intermediate-mass elements
Si was decreased and S increased, indicating that burning had been
much less effective in \bg\/ than in normal SNe~Ia. The resulting
synthetic spectrum had absolute magnitudes $V=-17.3$ and $B=-16.7$.
The new model, which includes photon cascades, can explain the re-emission
feature at 4500A, which was not reproduced with the previous pure
scattering model.

The near-maximum-light spectrum of \cn\/ was obtained on May 22, 1997,
i.e. 3 days after $B$ maximum. We therefore used $t=21$ days,
consistent with a rise time to $B$ maximum of 18 days, as for \bg.
Since the spectra of \cn\/ and \bg\/ are so similar, we assumed that
the temperature was the same in both, even though the epoch is
slightly different. We used the same luminosity as in the maximum
light \bg\/ model ($\log L / \Lsun = 42.33$ erg s$^{-1}$), an
assumption which is justified by the rather flat shape of the light
curve of \bg\/ around maximum (\cite{tur91bg}). We then rescaled the
photospheric velocity to a value that would yield the same $\Teff$ at
the new epoch, i.e. $v_{ph}=6000$ km s$^{-1}$.  Both the photospheric
velocity and the effective temperature are low with respect to the
values for a normal SN~Ia. We refer the reader to the paper discussing
modeling of \bg\/ (\cite{maz97}) for a more detailed discussion of the
effects of these quantities on the synthetic spectra.  We used a
distance $\mu = 33.95$, and reddening $E(B-V) = 0.0$.  The synthetic
spectrum thus computed reproduces well both the spectrum of \cn\/ at
$t=21$ days and that of \bg\/ at $t=18$ days (Fig.\ref{spmod}).

Although a more detailed description of the new MC code will be
presented elsewhere, we note that with this new version the amount of
\ion{Ti}{2} needed to model the spectrum of \bg\/ increases from $7
\times 10^{-5}$ (\cite{maz97}) to 0.005 \M, probably
because of the new atomic data. A small difference between the two
observed spectra (\cn\/ and \bg) is the depth and the shape of the
\ion{Ti}{2} trough between 4000 and 4500A, so that a mass of Ti of
0.002 \M\/ above the photosphere gives the best fit to the spectrum of \cn.

In the model, increasing only the luminosity results in a higher
photospheric temperature. However, for even a small increase of the
temperature the shape of the spectrum changes, and the fit becomes
unacceptable. Clearly, a good fit can only be obtained with a very
specific $\Teff$. In order to conserve that temperature, the
photospheric radius must be increased along with the luminosity.  If
we adopt a distance $\mu = 34.53$ , the luminosity increases by 0.58
mag to $\log L/\Lsun = 42.56$, and the velocity must increase from
6000 to 7750 km/s. Since the mass above the photosphere is smaller for
a model with higher $v_{ph}$ and a constant exploding mass of 0.7 \M,
(0.38 \M vs. 0.48 \M for the lower $v_{ph}$ model), the synthetic
lines are too shallow. This can be circumvented by adopting a higher
exploding mass. A model with the larger distance, the same input
parameters just listed and a mass of 1.4 \M\ gives a fit of the same
quality as the short distance, low velocity, small mass model. The
difference in the photospheric velocity is in fact too small to be
noticeable.

Current problems with the sub-Chandrasekhar explosion scenario (in
particular the predicted but not observed presence of an outer Ni
shell in current He-detonation models) may be used as an argument in
favour of the long distance scale. However, if we adopt the long
distance scale for \cn, and of course also for \bg, the same argument
should then be applied to normal SNe~Ia. This may lead to masses
significantly in excess of the Chandrasekhar limit, and so we would
only be shifting the problem from faint to average SN Ia models.

To be fair we can say that, because of the uncertainties in the
progenitor scenarios, spectral synthesis modeling of the photospheric
epoch alone are not sufficient to constrain the value of the Hubble
constant, as already stressed for \bg.  
Greater insight can be obtained using late-time spectra as
well, as was done for \bg. Unfortunately, \cn\/ was too faint to be
observable at such late epochs, but we expect that it would have had
narrow lines, as did \bg.

We have also modeled the light curve of \cn\/ using the simple Monte
Carlo code described in \cite{cap97}. We have compared the synthetic
bolometric light curve with the observed $V$ curve, assuming that the
bolometric correction is small. The mass of Ni was assumed to be
centrally concentrated. We find that for the short distance a Ni mass
of $\sim 0.1$ \M\ is required to reproduce the maximum magnitude. The
exact value of the ejecta mass depends on the kinetic energy (KE),
which depends on the production of Ni but also on that of Si. Values
of M$_{\rm ej}$ between 0.4 and 0.7 \M\ yield acceptable fits to the
light curve depending on whether the KE per unit mass is rescaled
according to the M$_{\rm Ni}$/M$_{\rm ej}$ ratio or not. In this
latter case we assume that burning to Si provides the additional KE to
make up for the small production of Ni.  

In Fig.~\ref{cdlmod} we show the fit to the observations of different
models.  In particular it appears that a good fit can be obtained for
a model with Mej = 0.4 \M, and KE rescaled according to the M$_{\rm
Ni}$/M$_{\rm ej}$ ratio. The assumption of no positron deposition
leads to a better fit at late phases, as was the case for
\bg. For the large distance a similarly good fit can be obtained for
M$_{\rm Ni}$=0.2 \M\ and M$_{\rm ej}=0.55$ -- 1 \M. Thus the fast
decline supports the sub-Chandra mass option at both distances.

In conclusion, good fits to the spectra and the light curve can be
obtained for both the short and the long distance scales.  However,
while in the case of short distances the ejecta masses indicated by
the light curve and the spectra agree rather well ($\sim 0.7$ \M),
this is not the case for the long distance, where the best fit
`spectroscopic mass' ($\sim 1.4$ \M) is larger than the `photometric
mass' ($\sim 1.0$ \M).

\section{Conclusion}

We have presented a set of observations of the recent \cn\/ which
demonstrate that it is a faint SN~Ia nearly identical to \bg\/. In
particular, we measured similar values for the light curve parameters
$\Delta m_{15}$ and the luminosity correction $\Delta$, which are
known to correlate with the luminosity. The two SNe are also similar
in that their light curves lack the secondary maxima at red
wavelengths typical of normal SNe~Ia, and they have similar colour
curves. The two objects share also the same spectral appearance, the
rapid evolution, and the low temperature and photospheric velocities.

The best fit model depends on the luminosity calibrations: if we adopt
a short distance scale we can fit the spectrum adopting a W7 density
profile scaled to a total mass of the exploding star of $\sim 0.7$ \M, as we
did for \bg. The mass of radioactive material synthesized during the
explosion, which is the main parameter characterizing SNe Ia
(\cite{cap97}), was 0.05--0.15 \M\ for \bg\/ (\cite{maz97}).  If on
the other hand we adopt a long distance scale a good fit can be
obtained with an unscaled W7 model.  Even in this case, though, the Ni
mass would be only $\sim 0.2$\M, which is small for a normal
Chandrasekhar mass SN~Ia.

In a recent paper (\cite{maz98}) we have shown that the luminosity
decline rate after maximum light, $\Delta m_{15}$, is correlated with
the expansion velocity of the Fe core measured from the width of the
nebular lines. Since $\Delta m_{15}$ is correlated with the absolute
magnitude at maximum, which in turn is directly related to the \ni\/
mass, this finding suggests that the amount of radioactive material
synthesized is directly correlated with the kinetic energy of the
ejecta. At low expansion velocities this relation is based on the
observations of only \bg. Unfortunately our attempt to recover \cn\/
in the nebular stage was unsuccessful, and we could only place an
upper limit which is however consistent with a behaviour similar to
that of \bg.

A common concern is that faint SN~Ia could be fairly frequent and
therefore severely underestimated in current searches. A recurrent
argument is that indeed faint SNe are only found in nearby galaxies.
We note however that modern CCD searches are more effective in
discovering faint, red objects like \bg\/ and \cn\/ than traditional
searches which use blue sensitive photographic plates.  The discovery
of \cn\/ in a relatively distant galaxy confirms that such searches
are indeed able discover faint SNe Ia in a fairly large volume.  Thus,
the question of the intrinsic frequency of faint SNe Ia will hopefully
be settled within the next few years.

It is also worth mentioning that the parent galaxy of SN~1997cn is
elliptical. In recent years evidence has been accumulating that SNe Ia
in ellipticals are on the average fainter that those in spirals.
\cn\/ is obviously consistent in this respect. Again, the statistics
still need to be improved, but the idea that the progenitors of SNe Ia
in ellipticals and in spirals may originate in different populations
merits further investigation.

\acknowledgements{We are pleased to thank the Leiden Observatory and
Mr. Brogt who kindly allowed the observations of \cn\/ at the Dutch
telescope also during the Dutch reserved time.}

\newpage

\figcaption[./fig/sn97cn_map.ps]{V band image taken with the Dutch 90cm 
 telescope on May 17 1997. The stars of the local sequence used for
 calibrating the SN photometry (cfr. Tab.~2) are
 identified. \label{snfield}}

\figcaption[./fig/sn97cn_obs.eps]{UBVRI photometry for SN 1997cn. The
continuous lines are the V templates of SN 1991bg and of normal SNIa
shifted on both axes to match the V maximum. The open symbols are
unfiltered CCD magnitudes from IAUC 6661. \label{figlc}}
 
\figcaption[./fig/sn97cn_cfr.eps]{Comparison of the absolute V and I
light curves and (B--V) colour curves of \cn\/ (filled symbols) with
those of \bg\/ (open circles), SN~1992K (open triangles) and SN~1994D
(open pentagons).  The similarity with \bg\/ both as to luminosity and
light curve shape is evident. Note the absence of the secondary
maximum in the I band and the early turnover of the (B--V) colour
curve. The distance moduli adopted are in the SBF--PNLF--TF distance
scale compatible with Ho=85 \kms Mpc$^{-1}$ ($\mu(97cn)=33.95$,
$\mu(91bg)=31.09$, $\mu(92K)=32.96$ and $\mu(94D)=30.68$). The
reddenings are $E(B-V)(97cn)=0.00$, $E(B-V)(91bg)=0.05$,
$E(B-V)(92K)=0.12$ and $E(B-V)(94D)=0.06$ are accounted
for. \label{complc}}

\figcaption[./fig/sn97cn_spcf.eps]{Comparison of the spectra of \cn\/
and \bg\/ in proximity of maximum and one month later. \label{compsp}}
                                    
\figcaption[./fig/sn97cn_spettro.eps]{The maximum light spectrum compared
to the model spectrum obtained for a W7 density structure scaled to a
total mass of 0.7 \M, $\log L=42.33$, photospheric velocity
$v_{ph}=6000$ \kms, rise time to B maximum is $t=21$ days. No reddening
has been adopted. Main line identifications are marked. \label{spmod}}
    
\figcaption[./fig/sn97cn_cdlmod.eps]{\label{cdlmod} 
Comparison of the $V$ light curve of
\cn\/ (triangles) with models. The dotted lines are the models
obtained with 0.1 \M\ of Ni and 0.40 \M\ of ejecta transformed to
apparent magnitudes with $\mu=33.95$ ($H_o=85$).  The short dashed
lines are models with 0.2 \M\ of Ni and 0.55 \M\ transformed to
apparent magnitudes with $\mu=34.53$ ($H_o=65$).  Models with two
different positron opacity $\tau_{e^+}$ are shown.  It has been
assumed that the bolometric correction is negligible and that
$E(B-V)=0.00$.}
	
\end{document}